# Thermal and electrical characterization of poly(vinyl)alcohol)/poly(vinylidene fluoride) blends reinforced with nano – graphene platelets

Eirini Kolonelou[a], Eirini Loupou[a], Panagiotis A. Klonos[b], Elias Sakellis[a,c], Dimitra Valadorou[a], Apostolos Kyritsis[b], Anthony N. Papathanassiou[a,*]

[a] National and Kapodistrian University of Athens, Physics Department, Section of Condensed Matter Physics, Panepistimioupolis, GR 15784 Zografos, Athens, Greece

[b] Department of Physics, National Technical University of Athens, Department of Applied Mathematics and Science, 15780 Zografos, Athens, Greece

[c] National Center of Natural Sciences Demokritos, Institute of Nanomaterials and Nanotechology, Aghia Paraskevi, Athens, Greece

[*]Corresponding author email: antpapa@phys.uoa.gr (A.N. Papathanassiou)

**Abstract:**

Nano-graphene / polymer composites can function as pressure induced electro-switches, at concentrations around their conductivity percolation threshold. Close to the critical point, the pressure dependence of the electron tunneling through the polymer barrier separating nano-graphene's results from the competition among shortening of the tunneling length and the increase of the polymer's polarizability. Such switching behavior was recently observed in polyvinyl alcohol (PVA) loaded with nano-graphene platelets (NGPs). In this work, PVA is blended with α-poly(vinylidene fluoride) (PVdF) and NGPs. Coaxial mechanical stress and electric field render the nano-composite piezoelectric. We investigate the influence of heterogeneity, thermal properties, phase transitions and kinetic processes occurring in the polymer matrix on the macroscopic electrical conductivity and interfacial polarization in casted specimens. Furthermore, the effect of electro-activity of PVdF grains on the electric and thermal properties are comparatively studied. Broadband Dielectric spectroscopy is employed to resolve and inspect electron transport and trapping with respect to thermal transitions and kinetic processes traced via Differential Scanning Calorimetry. The harmonic electric field applied during a BDS sweep induces volume modifications of the electro-active PVdF grains, while, electro-activity of PVdF grains can disturb the internal electric field that free (or bound) electric. The dc conductivity and dielectric relaxation was found to exhibit weak dependencies.

# 1. Introduction

Polyvinyl alcohol (PVA) is a water-soluble synthetic polymer, which has excellent film forming, emulsifying and adhesive properties [1]. It has high tensile strength and flexibility which is comparable to those of human skin and tissues. Its hydroxyl group favors hydrogen bonding and, hence, makes it biocompatible. [2] Polyvinylidene fluoride (PVdF) is a well-known for its premium properties, such as high electrical insulation, remarkable thermal and chemical stability, molding efficiency, and biocompatibility [3]. Its strong polar nature has its roots to the significant dipole moment of the fluorine – carbon bond. The most common crystalline phase, known as the α-phase, is non-piezoelectric, which can transform to one of its piezoelectric polymorphs: β, γ or δ-phases. The intense electro-activity of these phases also makes them exceptional organic materials for energy conversion and spare energy harvesting [4, 5]. Currently, researchers focus on PVdF based flexible nano- and micro-patterned polymer surfaces for biomedical sensing and energy harvesting applications [6]. The α -phase has semi-helical TGTG′ (trans-gauche–trans-gauche) conformation and is of non-polar nature due to the anti-parallel packing of dipoles inside the unit cell, resulting in no electro-active properties. In case of β (all Trans TTT chain conformations) and γ (T3GT3G′ conformations) phases, all dipoles of individual molecules are arranged parallel to each other producing a non-zero dipole moment and induce polarity. Also, PVdF and its copolymers can be dissolved in liquid solvents so that they can be extruded or molded into desired shapes or directly coated onto a substrate, adding more flexibility in transducer design [7]. PVdF has been found suitable for a variety of energy-related applications such as ultra-transducers, audio transducers, medical transducers, display devices, vibrometers, shock sensors, pressure sensors. [7, 8].

Graphene nanoparticles in PVA enhance the properties of the composite, such as electrical conductivity and thermal stability [9]. Nano-graphene platelets (NGPs) is an affordable choice to reinforce a polymer. At mass fractions low enough to prevent a continuous [percolation network due to physical contacting among NGPs, the composite's electrical conductivity can be comparable to that of a semiconductor. It was found that the percolation threshold in polymer nanocomposites (PNCs) loaded with various carbon allotropes [10, 11, 12] is achieved for quite low mass fractions. Thermal fluctuations of the Fermi level of electron states in NGPs, induce inter-NGP tunneling through the separating polymer barrier. We distinguish the difference between the *physical network* of NGPs formed at sufficiently high concentrations of NGP the polymer nano-composite (PNC) and the *electron percolation network*; the latter consists of *isolated* NGPs ( or isolated aggregates) and the inter-NGP polymer regions through which the inter-NGP tunneling current flows. The tunneling current depends on the tunneling volume or the inter-NGP distance, the polarizability of the polymer matrix and its thermodynamic state (glassy, semi-crystalline or rubber). By tuning the pressure dependencies of these factors, it is possible to develop pressure sensing or pressure electro-switching PNCs. Hence, it is essential to resolve particular electric charge flow mechanisms and relate them with heterogeneity, the glassy or rubber state of the polymer and kinetic phenomena that may occur, such as the mobility of solvent water molecules. The phenomenon of pressure induced electro-switching has been observed in PVA/NGP composites, stemming from the competition among the above-mentioned factors [12.

In the present work, we prepared NGP composites in PVA blended with 25 wt. % α-poly (vinylidene fluoride) (PVdF) micro-grains. The structural and thermal stability of PVA/PVdF blends, as well as, their structural and thermal properties and the interaction among their different phases have been investigated by using XRD, FTIR and Raman spectroscopy and consular mass fraction of 22.5 wt. % PVdF are ideal to be used as a polymer matrix for electrical applications [13, 14]. The composite becomes piezoelectric, after mechanical stressing and polarization. Combined thermal and dielectric experiments aim to reveal the role of heterogeneity, structure, kinetic effects, phase transitions and (induced) electro-activity of PVdF on electron charge transport. Concerning the role of electro-active PVdF grains, we turn the authors; attention to the fact that the harmonic electric field applied during a BDS sweep induces – in principle - volume modifications of the electro-active PVdF grains and, subsequently, a time-varying stress field develops within the blend. On the other hand, electro-activity of PVdF grains may disturb the internal electric field that free (or bound) electric charges sense. The experimental methods used are: (i): Scanning Electron Microscopy (SEM) to profile the heterogeneity and check the NGPs dispersion, (ii): Differential Scanning Calorimetry (DSC) for the thermal transitions characterization, with emphasis on the glass transition and melting as well as to the evaporation process of water which inherently remains into the bulk in equilibrium with ambient humidity, (iii): time-domain electro-mechanical coupling to test the electro-mechanical functionality of fresh samples processed by axial compression and polarization to achieve an overall piezoelectric character, (iv):Broadband Dielectric Spectroscopy (BDS) to unravel and identify the origin of different dc conductivity mechanisms and dielectric loss.

## 2. Experimental

### 2.1 Blending procedure

The starting materials used for the preparation of the bulk specimens comprised polyvinyl alcohol (PVA, $(C_2H_4O)_n$, purity 99,5% ; ASG SCIENTIFIC, CAS 9002-89-5), poly(vinylidene fluoride) (PVdF, Mw 534.000, powder ( Sigma Aldrich Ltd) CAS 24937-79-9) of mean grain size 2.5 μm [15] and graphene nano-platelets (NGPs) (Angstrom Materials Ltd) which were used without any further purification. According to the manufacturer, the lateral size of NGPs are ≤ 10 $μm^2$ and the average through dimensions were 50 – 100 nm determined by BET surface analysis and size distribution. Earlier publications on water soluble polymers (such as PVA, PVP and blends) ) reinforced with different mass fractions of NGPs by solution casting similar to that mentioned in the present work, indicated that the critical mass fraction of NGPs ranges from 0.1 to 0.3 % wt. % [11, 12, 13]; the conductivity percolation cluster is formed at much lower fraction than that required for the formation of a continuous physical NGPs, due to the inter-NGP tunneling current through the polymer barrier. Here, the mass fraction is 2 wt. %, i.e., at a value close enough and above percolation threshold and, at the same time, low enough to prevent the formation of a continuous network of NGPs.

**Table 1**

| sample description | code name | $h_{amb}$ (wt.%) | $T_g$ ($^oC$) (±5 $^oC$) |
|---|---|---|---|
| PVA 75% - PVdF 25% (unfilled matrix) | Matrix | 0.03 | 20 |
| Nano-composite with 2% NGP | PNC | 0.13 | 20 |
| Nano-composite with 2% NGP + Piezo-electric | PNC-PZ | 0.08 | 27 |
| PVA neat | PVA | 0.05 | 50 |
| PVdF neat | PVdF | 0 | - |

***Table 1.*** *Labeling and description of the materials under investigation, ambient water content, $h_{amb}$, as estimated by comparing the wet and dry sample mass, and the glass transition temperature $T_g$,.*

PVA powder was dissolved in doubly distilled and de-ionized water at 60$^o$ C by continuous stirring for about 1 h to form an aqueous solution of 9 wt.% PVA. Earlier tests in our laboratory proved that the latter concentration offers suitable viscosity to ensure optimum homogeneous dispersion od NGPs and PVdF powder, as verified by SEM (Scanning Electron Microscope) inspection of the end solid composites. Similarly, NPG powder (~0.4gr) was dispersed in water (~2ml) to form a suspension of 2wt % NGP, by continuous action for about 30 min, in order to break clusters of NGPs. Afterwards, the NGP suspension was added in the PVA solutions and continuously stirred, while, PVdF powder was added. Subsequently, the solution was stirred and placed in an ultrasonic bath for about 1 h. The product (~5g) was poured onto a flat teflon surface to form a layer of area ~50 cm$^2$). Water was allowed to evaporate at ambient conditions, i.e., room temperature 22-24 $^o$C and relative air humidity 50-55 % for 48 h, yielding free-standing solid samples of thickness about 70 μm.

*2.2 Formation of piezoelectric composites from a fresh one: experimental procedure and electro-mechanical coupling characterization*

In the present work, the non-piezoelectric blends are labeled "fresh". A simple and effective strategy applied in a recent publication of ours [16, 17] to render freestanding disc-shaped fresh PVdF composites: A stress of 1.3 GPa applied normal to the parallel surfaces of the disc and, subsequently, an electric field of intensity 30 MV/m applied along the direction of the applied stress, for 24 h, at about 315 K. Finally, the electro-activity of the blend was inspected as follows. The open circuit potential difference between the parallel surfaces of the disc-shape specimen was recorded by using a Keithley 617 Electrometer, while a mechanical stimulus was applied to the parallel surfaces.

*2.3 Characterization/ techniques*

The micro-morphology of the samples was observed by SEM (FEI inspect microscope, USA) operating at 25kV, equipped with an EDAX super ultra-thin window analyzer for energy dispersive X-ray spectroscopy (EDS). The thermal transitions of the polymers, with a focus on glass transition and melting of crystals, were investigated in nitrogen atmosphere (99.9995% purity) in the temperature range from −100 to 230 $^o$C by means of a TA Q200 series DSC instrument (TA Instruments, USA), calibrated with Indium for temperature and enthalpy and sapphires for the heat capacity. The samples were measured as received, moreover, pieces of the prepared films of ~6 mg in mass were closed in standard aluminium TA pans. The heating rate was fixed at 10 $^o$C/min for all compositions.

Isothermal dielectric measurements (BDS) were carried out using the sample cell of Novocontrol pressure – temperature apparatus, kept at ambient pressure, connected to a Solartron SI 1260 Frequency Response Analyzer covering a frequency range from $10^{-2}$ to $10^6$ Hz. The temperature stability was better than 0.1$^o$C. Samples were placed between two bronze flat electrodes, thus forming a metal-dielectric-metal structure (capacitor). Quick drying silver paste was used to ensure good electrical contacts. BDS measurements were performed at ambient pressure and temperatures ranging from 293 K to 393 K at steps of 10 K.

## 3. Results and Discussion

### *3.1 SEM imaging*

Figure 1 shows the BSE (Backscatter electron) images obtained for the PNCs, whereas NGPs seem to be distributed homogeneously. By gradually increasing the high voltage, information was gained from larger regions extending from the surface towards the interior and better visualization of NGP flakes. Typical dimensions < 4 μm could be measured for the NGPs. The latter indicates that the maximum size of their large planes can hardly exceed that reported by the manufacturer and, thus, evidence for the absence of large NGP agglomerates. In Figure 2, EDS spectra and quantitative analyses, collected from the light grey and dark grey phases appearing in the SEM images (Figure 1). The quantitative analyses for C, O and F elements and, especially, the relative abidance of F, can attribute the light grey regions to PVdF and the dark grey one to PVA. Inspection of the SEM images evidence about phase separation between the components of the polymer matrix.

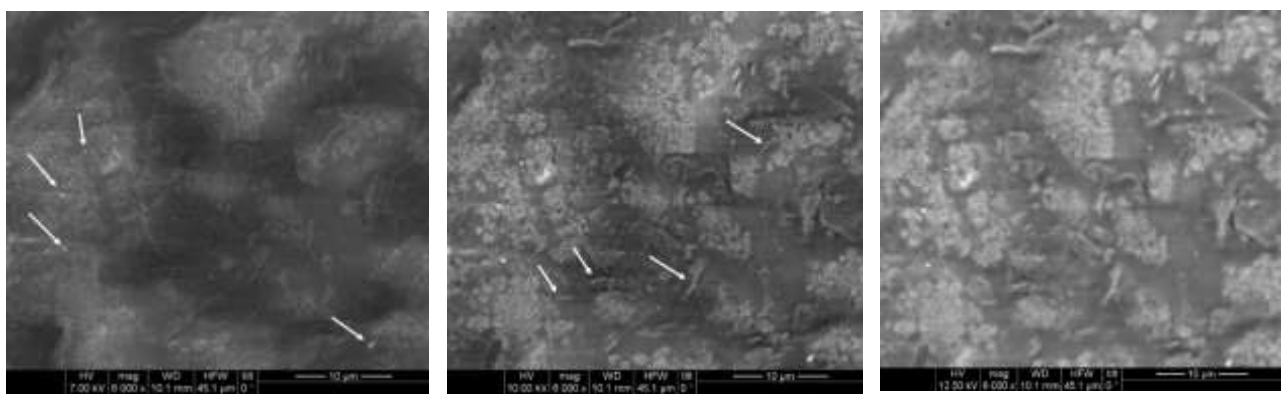

*Figure 1. BSE (Backscatter electron) images of the PNC blends, with gradually increasing electron accelerating voltage (from left to right) and, hence, sampling regions extending from the surface to gradually increasing depth. As explained in the text, dark grey regions are identified as dominated by PVA, while light grey ones PVdF. The white arrows in the middle photo, point towards NGP flakes.*

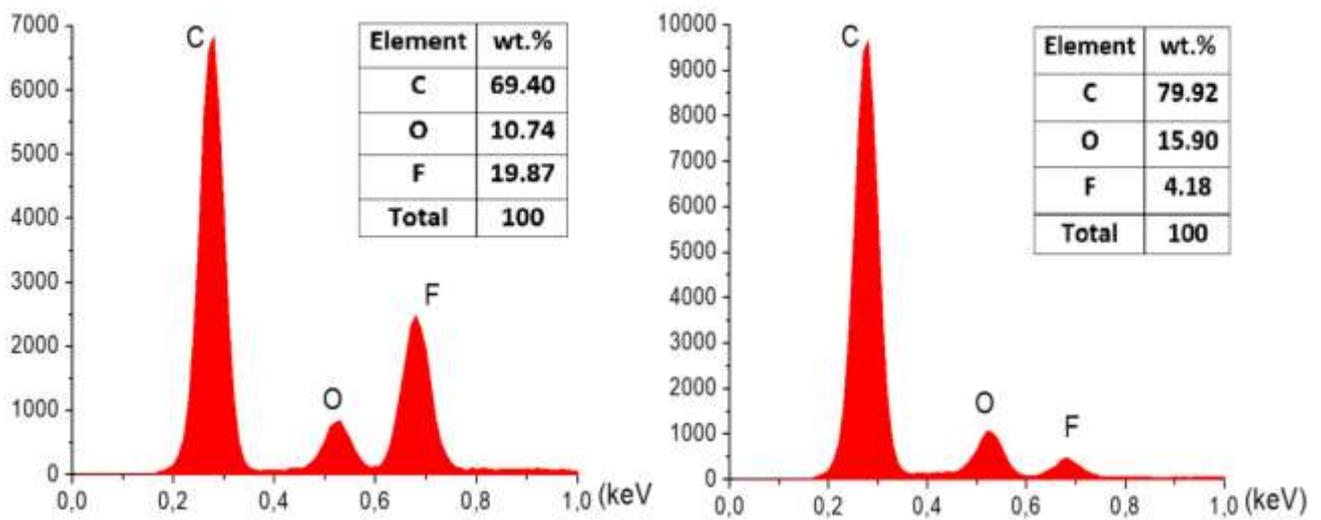

*Figure 2. EDS spectrum of C, O and F collected from light and dark grey regions of the SEM images, depicted in the left and right diagram, respectively. (Vertical axes are in a.u.) The intense fluorine peak signs PVdF, rather than PVA. The quantitative analyses are depicted as inset tables.*

### 3.2 Time – domain electro-mechanical coupling

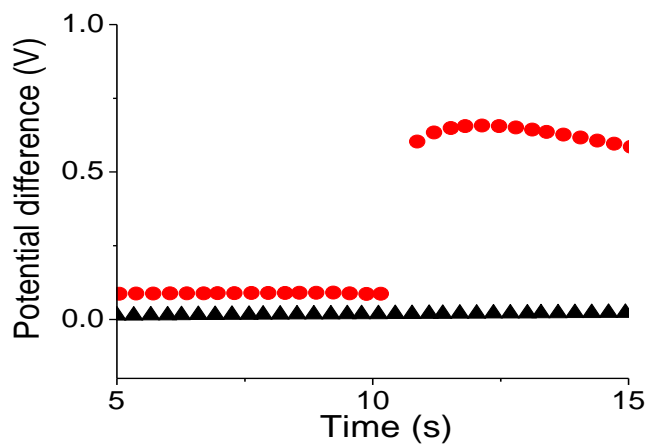

*Figure 3. The open circuit potential difference measured between the two parallel surfaces of a disk-shaped piezoelectric specimen, as a function of time. A step-like force of amplitude F = 0.01 N was exerted normal its parallel (circles). Data points collected from its fresh (non-piezoelectric) blend are also depicted (triangles).*

Piezoelectricity in specimens processed through combined mechanical deformation and electric polarization, was tested by measuring the open circuit voltage developed among the two opposing surfaces of a disk-shape composite when an external force is applied (Figure 3). If *j* denotes the common direction along which mechanical stimulus is applied and voltage V is measured, the (absolute) value of the piezoelectric coefficient was obtained using the formula: $d_{jj} = CV/F$, where F is the magnitude of the force and C the capacitance.. The capacitance was measured directly by using a high precision Solartron 1260 Frequency Response Analyzer. The electro-mechanical coupling was estimated: $d_{jj} \simeq 10 \; pC/N$

## *3.3 Differential Scanning Calorimetry (DSC)*

The samples were measured by DSC as received in order to make a direct comparison with results by BDS. Figure 4 shows the respective DSC heating thermograms for PNC, PNC-PZ and the unfilled PVA-PVdF matrix (please compare with Table 1). Since many complex phenomena contribute to the shown heating thermograms, DSC measurements were carried out also in neat PVA and neat PVdF, in order to clarify the thermal transitions recorded in the PNCs.

In Figure 4, a step-like endothermic event is recorded at ~293 K (~20 $^o$C). Comparing with data for neat PVA in figure 5 and, taking into account the large fraction of PVA in the PNCs, it is likely that this step-like event corresponds to the glass transition of PVA. A rough estimation of the glass transition temperature, $T_g$, was performed from the derivative of heat flow. Accompanied by a significantly large uncertainty (≤278 K (5 $^o$C)), $T_g$ was found as ~293 K (~20 $^o$C in Figure 4) in the matrix and the PNC, higher ~300 K (~27 $^o$C in Figure 4) in the PNC-PZ, whereas as for the more clear step of neat PVA $T_g$ equals ~323 K (~50 $^o$C in Figure 5).

The step of glass transition is followed by a strong endothermic peak, exhibiting maxima between 343 and 373 K (70 and 100 $^o$C in Figure 5). The peak is related with the evaporation of water that is expected in the samples, due to the hydrophilic character of PVA. Additional evaporation-related peaks are recorded for all samples (with the exception of neat PVdF) at temperatures between 443 and 473 K (170 and 200 $^o$C in Figure 5), these being most probably related with the evaporation of more bound water molecules from PVA. Since the used DSC pans were not sealed, the evaporated water was removed upon the heating scan. In a second heating scan performed subsequently (not shown), the evaporation peaks of both types vanished. Upon comparing the sample mass prior (hydrated) and upon ('dried') evaporation we made an estimation of the ambient water content, $h_{amb}$. The $h_{amb}$ values have been added in Figures 4 and 5 and in Table 1. Interestingly, while $h_{amb}$ equals ~5% and ~3% in the PVA-PVdF matrix, quite larger amounts were estimated for the two nano-composites, namely ~13% and ~8% for PNC and PNC-PZ, respectively. Also, upon the mentioned second heating, clear glass transition steps were recorded for PVA and all PVA-containing samples in the range between 333 and 353 K (i.e., 60 and 80 °C, respectively, as can be seen in Figure 5).

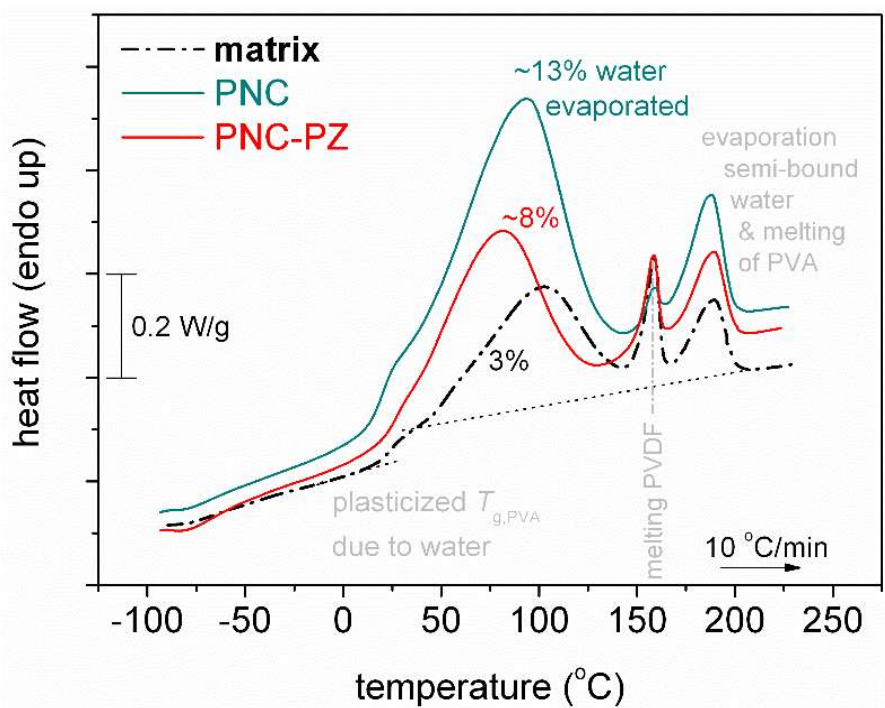

*Figure 4. Comparative DSC heating thermograms for the unfilled PVA-PVdF matrix, PNC and PNC-PZ as received. The heat flow data have been normalized to the sample mass.*

Combing these findings together, we conclude that the glass transition-like step observed in the first scan at the lower temperatures of ~293-323 K (~20-50 °C) arises from the hydrated PVA. The effect of lowering of the $T_g$ value upon polymer hydration is well known as 'plasticization' [18]. It is, however, interesting that the amount of hydration water has not systematically altered the $T_g$. This is actually a first indication that more parameters affect polymer mobility, for example crystallinity and the presence of NGP fillers.

Finally, at ~433 K (~160 °C) in Figure 4, single peaks are recorded for all samples containing PVdF, i.e. including neat PVdF in Figure 5. The position of the peak is identical for all samples and should arise from the melting of PVdF crystals. Regarding the glass transition of PVdF, with the exception of neat PVdF in Figure 4 that exhibits an extremely weak event at about ~313 K (~40 °C in Figure 5), we have no clear recordings on the segmental mobility of PVdF in the matrix and the PNCs.

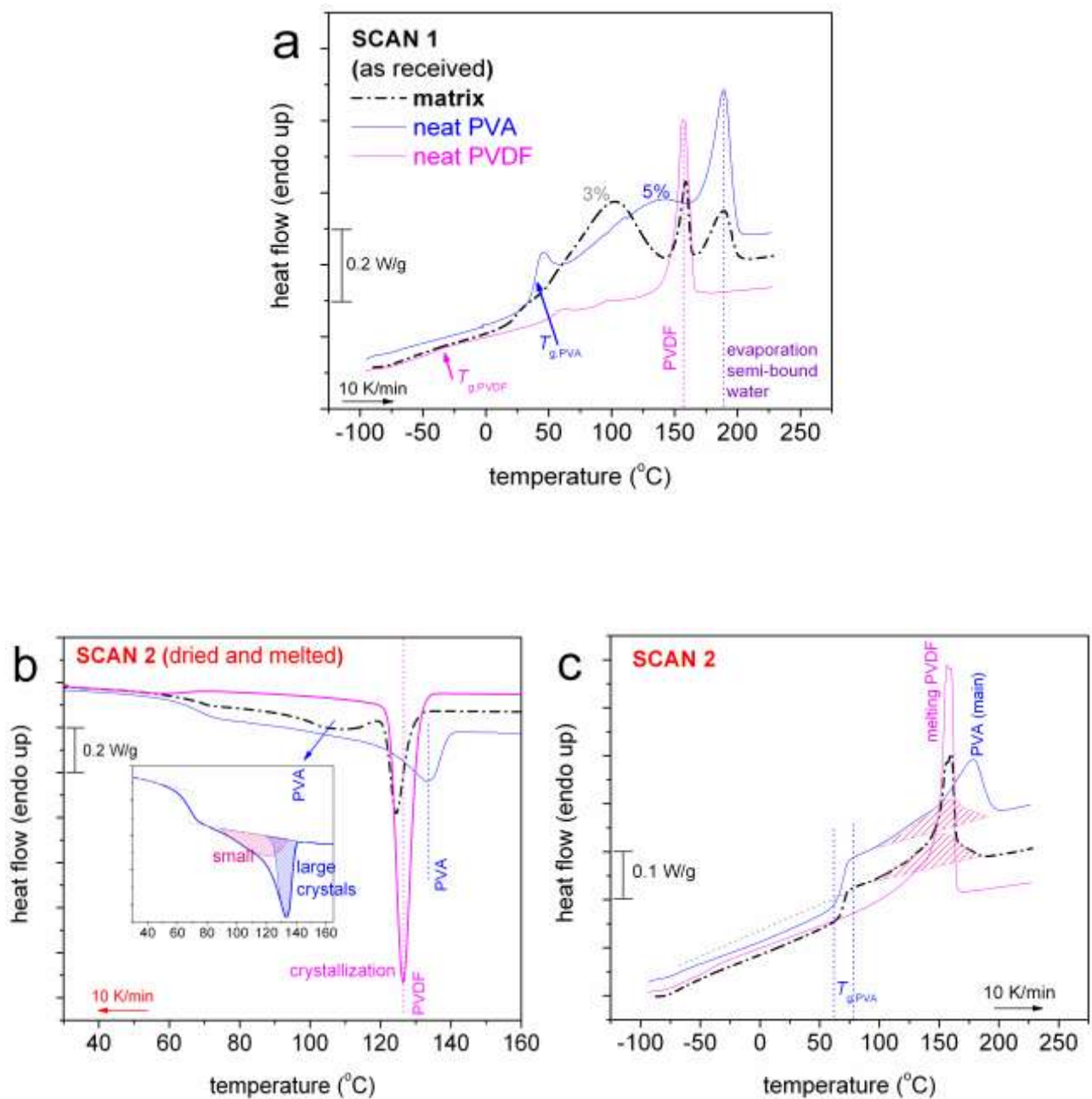

*Figure 5.* Comparative DSC thermograms for all samples during (a) scan 1 – samples as received and (b, c) scan 2 – cooling/heating. The heat flow data have been normalized to the sample mass. The heat flow data have been normalized to the sample mass.

Recording individually in the blend-matrix and the PNCs, the thermal events related to each polymer, suggests that there is a serious phase separation between PVA and PVdF, in AGREEMENT with the SEM profiling, where, as mentioned above, phase separation among the components of the blend is observed, The latter is wanted for these materials, envisaging their applications. More work could be made to further evaluate the data by DSC, namely in terms of crystalline fraction, alternations in the semi crystalline morphology, moreover, to estimate the mobile and rigid amorphous fractions of the polymers in the blend and in PNCs. more specialized techniques and measurement protocols should be employed, for example DSC measurements of the conventional as well as the temperature modulation mode using also sealed pans to keep hydration within the samples [19, 20]. Such work in the future, could provide more clues about the direct and indirect effects of polymer blending and filler addition on the polymer crystallization (nucleation, rate, and fraction) and segmental mobility.

*3.4 Dielectric Spectroscopy*

**3.4.1 Multi-scale electric charge transport and its temperature dependence in PNC with dispersed conductive NGPs**

Fluctuation Induced Tunneling (FIT) theory, founded by Sheng [21], is commonly used to interpret electronic transport in heterogeneous materials consisting of an insulating matrix and dispersed conducting inclusions [21]. Initially, FIT explained electronic transport in granular metals, whereas inter-grain tunneling of electrons through a separating insulating barrier was assisted by thermal fluctuations of the Fermi energy level. Furthermore, it was successfully applied for many different electron-conducting in homogeneously disordered materials, such as conducting polymers and composites. Accordingly, the temperature dependence of the dc conductivity $\sigma_{dc}$ is:

$$\sigma_{dc} = \sigma_0 exp\left(-\frac{T_1}{T+T_0}\right) \quad (1)$$

Sheng and Klafter [18,19] proved that the latter function reduced to a $T^{-1/2}$ dependent one:

$$\sigma_{dc} = \sigma_0' \exp\left(\frac{T_1}{T}\right)^{1/2} \quad (2)$$

Both (macroscopic) dc conductivity and localized charge motion stem from a unique type of elementary charge flow events in atomic scale. If $\Gamma_{ij}$ denotes the transition rate of electric charges tunneling from site i to site j, under the influence of an external field, then, $\sigma_{dc} \propto \Gamma_{ij}$ and $\tau^{-1} \propto \Gamma_{ij}$, where $\tau$ is the dielectric relaxation time. In the frequency domain, dielectric loss maximizes at $f_{max} = 1/\tau$, which implies that $f_{max} \propto \Gamma_{ij}$. Hence, the temperature dependence of the relaxation maxima resembles that of $\sigma_{dc}$, i.e.:

$$f_{max} \propto exp\left(-\frac{T_1}{T+T_0}\right) \tag{3}$$

and approximately:

$$f_{max} \propto \exp\left(\frac{T_1}{T}\right)^{1/2} \tag{4}$$

We stress that, although the approximate $T^{-1/2}$ dependence is predicted both by FIT and Mott's Variable Range Hopping (VRH), the underlying physics is different: in brief, FIT accounts for in homogeneously disordered materials whereas electron states extended over the volume of each one conducting grain and tunnel to neighboring grains due to thermal fluctuations of the Fermi level, while, VRH occurs in homogeneously disordered materials (i.e., amorphous semiconductors) by phonon assisted tunneling of electrons from a localized state to another one.

*3.4.2 Complex permittivity and tanδ*

Complex permittivity $\varepsilon^*(f) = \varepsilon'(f) + i\varepsilon''(f)$, where $f$ is the frequency of an externally applied harmonic field $f=\omega/(2\pi)$, $\omega$ being its angular frequency) and $i^2 = -1$, can capture both dc-conductivity and dielectric relaxation phenomena. The imaginary part $\varepsilon''$ consists of a dc-conductivity term $\frac{\sigma_{dc}}{\varepsilon_0 2\pi i f^n}$ and a sum over $m$ Harvilliak-Negami (HN) relaxation peaks $\sum_{m=1}^{k} \frac{\Delta\varepsilon}{\left(1+(f/f_{0,m})^{a_m}\right)^{b_m}}$, i.e.:

$$\varepsilon^*(f) = \frac{\sigma_{dc}}{\varepsilon_0 2\pi i f^n} + \sum_{m=1}^{k} \frac{\Delta\varepsilon}{\left(1+i(f/f_{0,m})^{a_m}\right)^{b_m}} \left(\frac{f}{f_{0,m}}\right) \tag{5}$$

where $\sigma_{dc}$ denotes the dc-conductivity, $n$ is a fractional exponent ($0 < n \leq 1$), $\Delta\varepsilon_m$ is the strength of $m$-th relaxation mechanism, $a_m$ and $b_m$ are its broadening and asymmetry shape parameters, respectively, and $f_{0,m}$ is a parameter that coincides with the peak maximum frequency $f_{max,m}$ of the $m$-th relaxation, provided that $b_m=1$. In Figure 6, it is hard to define dielectric relaxation peaks, due to the strong dc conductivity component. Relaxations with well-defined maxima can be seen in the tangent of the loss angle

$$tan\delta(f) = \frac{\varepsilon''}{\varepsilon'} \,. \tag{6}$$

representation (Figure 6). The peak maximum frequency $f_{max,\ tan\delta}$ have practically the same temperature dependence as the loss maxima in the ε´(f) domain.

*3.4.2 Dielectric relaxation mechanisms*

DSC measurements reveal the existence of absorbed water in the systems and that its evaporation starts already at ~ 50 °C. Traces of absorbed water remain in the volume of the free standing blend after the water solution casting at ambient conditions (of temperature, pressure and humidity). Systematic weighting of the fresh solid specimens for one week indicated that, the composite accommodates a practically constant concentration of water molecules, within the first 48 h of the casting. Evidently, the quantity of water absorbed in the bulk material is in equilibrium with the at the ambient conditions of our laboratory mentioned above. *We stress that the blends studied in the present work are scheduled to function in sensor and switching devices at environmental conditions and, for this reason,, it is desirable to confine the inherent water molecules, as long as dielectric spectroscopy experiments are conducted.* For this reason, in the BDS experiments, the entire capacitor holding the specimen was sealed with a thin flexible film of teflon and placed inside the Novocontrol High Pressure vessel, kept at ambient pressure. Hence, adsorbed water molecules could not escape from the volume of the specimen between the two parallel silver pasted surfaces. Finally, we checked if the possibility of water loss after a full set of BDS scans at gradually increasing temperatures, but, within the experimental accuracy we could achieve, no change of the weight of the sample was observed

BDS is a powerful tool to interrogate polymeric material characteristics because both the conductivity and chain motion can be monitored in the same spectra. In practice, electrode polarization effects interfere with the bulk response. Alternatively, researchers suggested using complex electric modulus and conductivity functions to analyze the dielectric spectra of the conductive materials. These functions are related via the equation: $tan\delta(f) = \frac{\varepsilon''}{\varepsilon'}$ [22-24].

The dielectric relaxation map can be visualized better by employing the *tanδ* function. In Figure 6, the dielectric loss factor (tanδ) had a maximum value. This maximum shifted toward higher frequencies as the temperature increased. These maxima may indicate the presence of interfacial polarization and the shift of the tanδ maxima toward higher frequencies as the temperature increases. These changes may be attributed to the existence of two different polarization mechanisms.

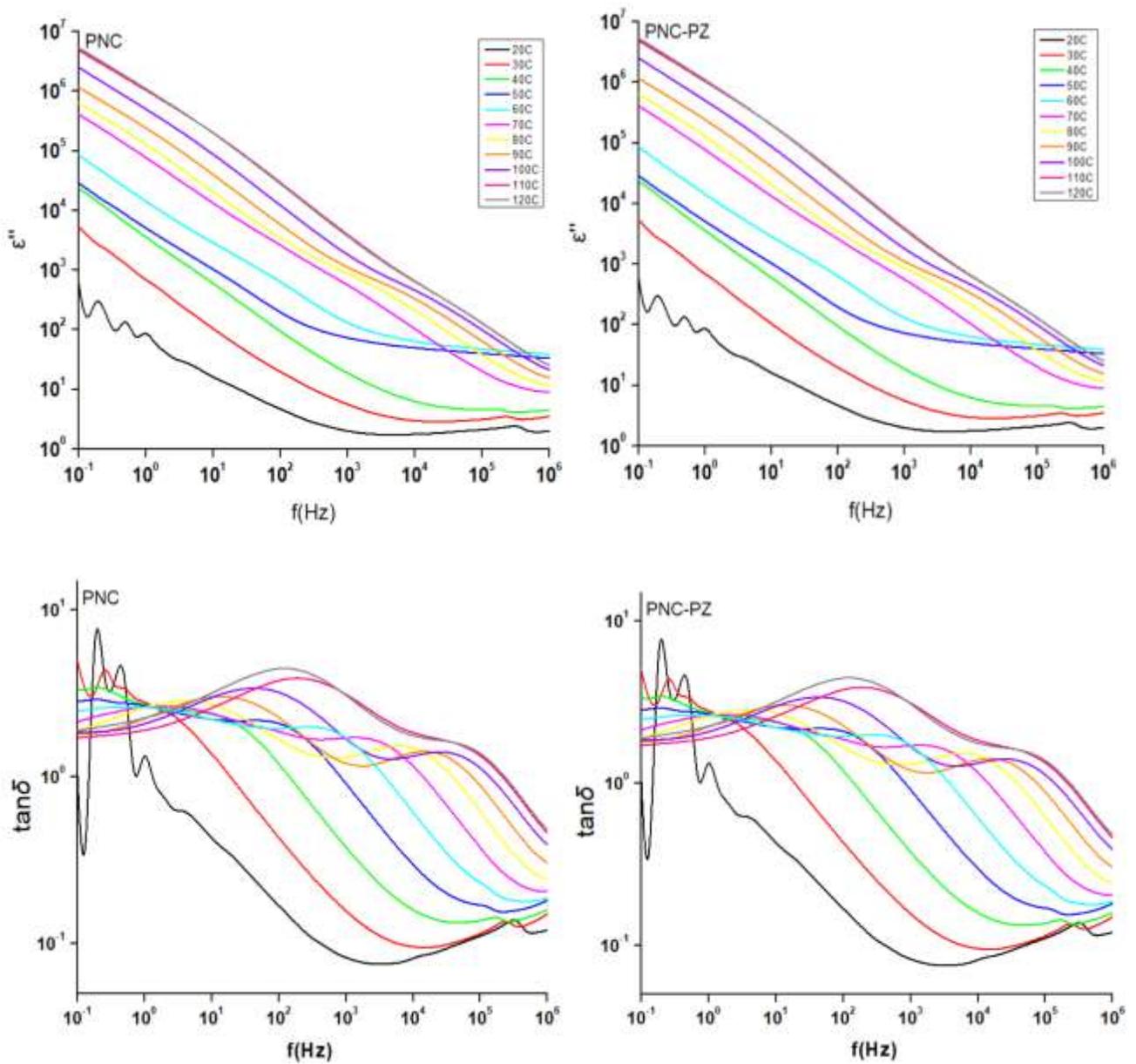

***Figure 6.*** *ε″ and tanδ spectra at different temperatures for PNC and PNC-PZ.*

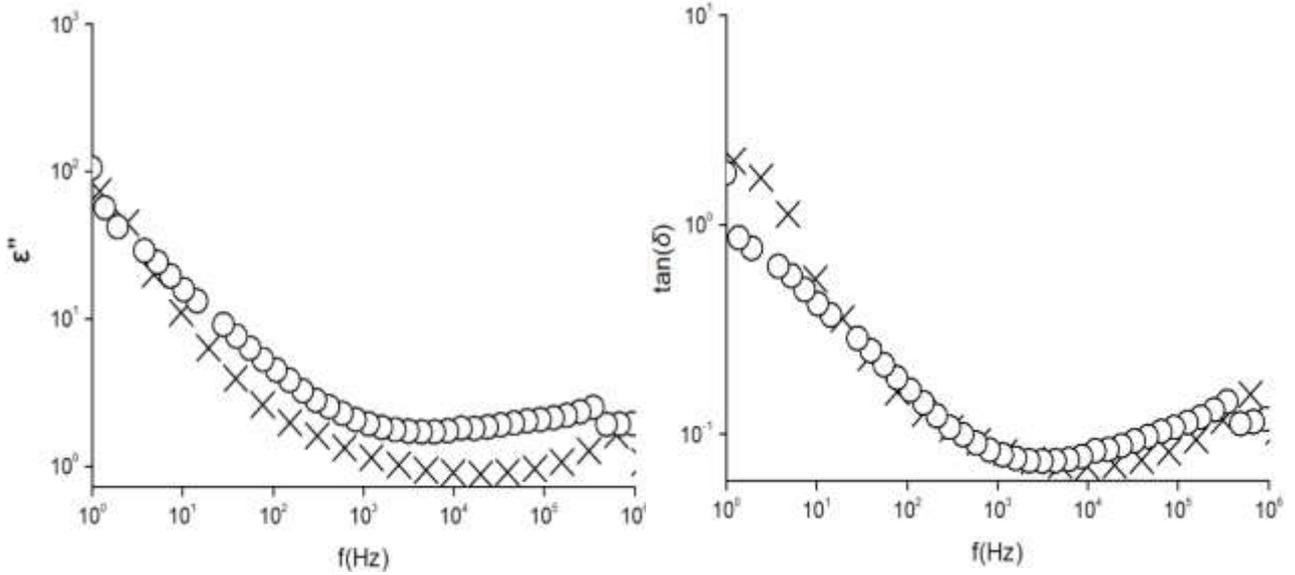

*Figure 7. Comparative measurements on an as-received sample (circles) and another one which was dried at 373 K for 24 h (cross) resulting in a reduction of its mass by 4 wt.%, due to the extraction of water molecules.*

As an example, the real and imaginary parts of the relative permittivity and the loss tangent obtained from impedance data were plotted as a function of frequency for the sample PNC-PZ (Figure 6). The mass fraction 2 wt. % NGP is higher than the percolation threshold, thus, inter-NGP electron tunneling dominates over any ionic or proton transport or molecular motion. Indeed, in Figure 7, the dielectric spectra of an as received sample and another one dried at 373 K for 24 h, yielding to 4 wt. % loss of mass due to the extraction of absorbed water molecules.

*tanδ(f)* isotherms are shown in figure 6c. Each curve observed consists of a dc conductivity line overlapping with one or more HN relaxation peaks (the HN function has initially been proposed as to analyze ε´´(f) results, since it is an empirical one and we are merely interested in the maximum frequency, we it may apply to match the *tanδ(f)* data points, as well).. A typical analysis of the *tanδ(f)* spectra is depicted in figure 7, where the experimental data points are fitted by a couple of HN peaks, namely DR1_tanδ and DR2_tanδ, respectively.

We attribute the relaxations to localized electron transport dictated by the electrical heterogeneity of the PNCs, for: (i) reinforcement with NGPs results in the appearance of the relaxations, (ii) $logf_{max, tan\delta}(T^{-1/2})$ obey a FIT model given by eq. (4a) (Figure 9), (iii) the $logf_{max, tan\delta}$ r temperature dependence evidence for low activation energy values, of the order of a few *meV*, which is typical for phonon assisted or fluctuation induced tunneling.

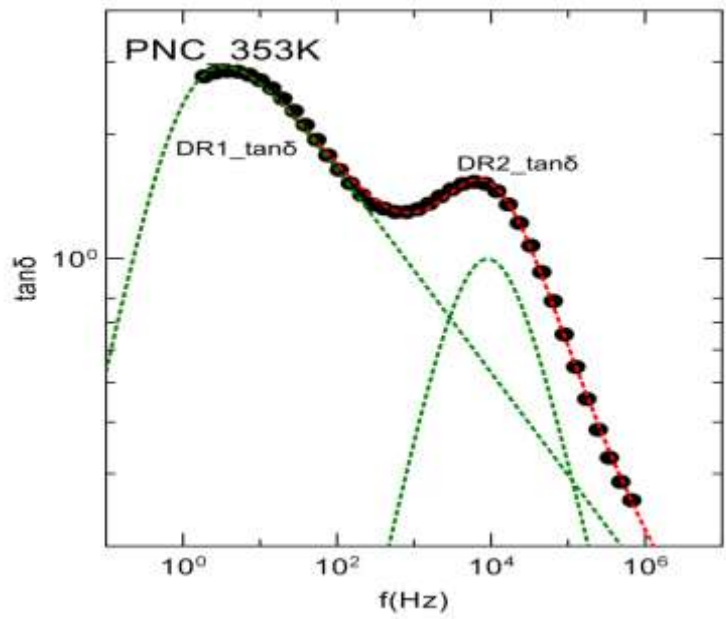

*Figure 8. Full line is the best fit to the data points collected at 353 K, consisting of two individual HN relaxation peaks (dash curves).*

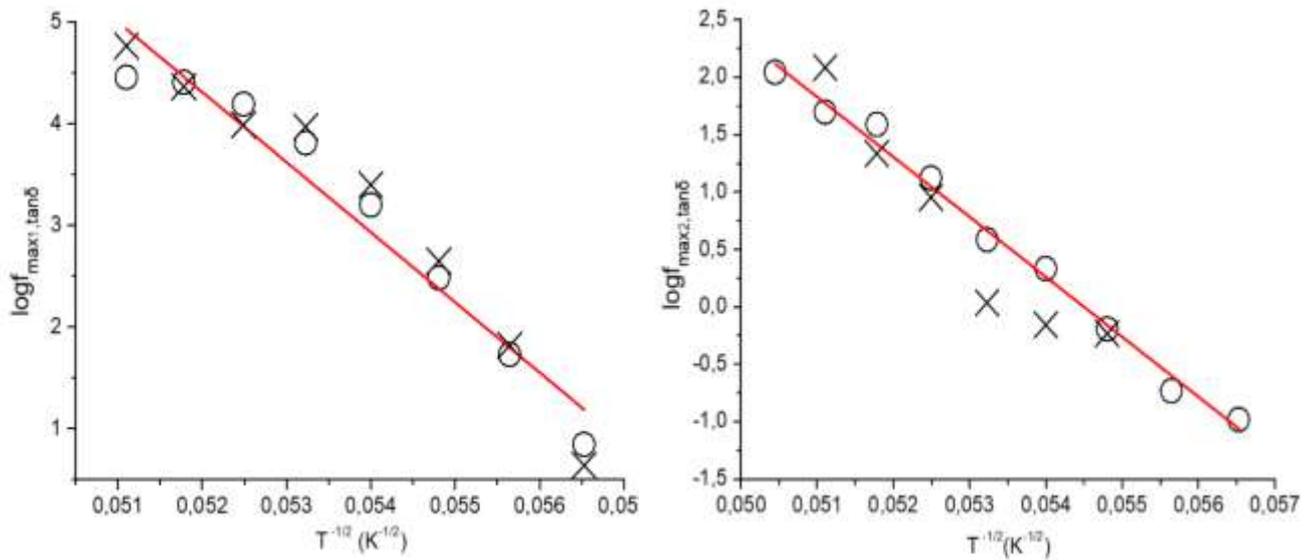

*Figure 9. Straight lines are best fitted, according to eq. (4), to $\log f_{max,\tan\delta}(T^{1/2})$ data for DR1_tanδ and DR2_tanδ, respectively. Circles correspond to fresh sample and cross to the piezo-electric nano-composite, respectively.*

In Figure 9, the entire set of data points fitted by a unique straight line. This means that relaxation of tunneling electrons, which is a local event, does is insensitive to global thermal events occurring in the polymer matrix; such regions could likely be NGPs themselves or NGPs and their immediate surroundings. Fitting eq. (3) to the $logf_{max,tan\delta}(T)$ data points (Figure 10), the activation energy, which is roughly equal to the related effect potential barrier, can be estimated from the parameter $T_1$ of the FIT model (eq (4)) [18]: $E \approx kT_1$, where $k$ is the Boltzmann constant. Their values are: 59,4 meV and 44,9 meV, for DR1_tanδ and DR2_tanδ, respectively. The presence of two relaxation mechanisms is probably related to the anisotropy of electron conduction within NGPs; the low activation energy mechanism (DR1_tanδ) is likely to capture on plane charge transport in grapheme layers, while, the high energy one (DR2_tanδ) to off-plane. Within an experimental error of about 15 %, each one of these values are identical, either for PNC or PNC-PZ, supporting the aforementioned scenario of relaxation occurring within NGPs, as follows: PNCs render piezoelectric by plastic deformation of a fresh specimen which modifies the shape of regions or grains of each polymer component, while the volumes of individual NGP remains unaffected; the energy landscape within NGPs remains unaffected. Thus, the activation energy for relaxation is independent upon the plastic deformation of the specimen.

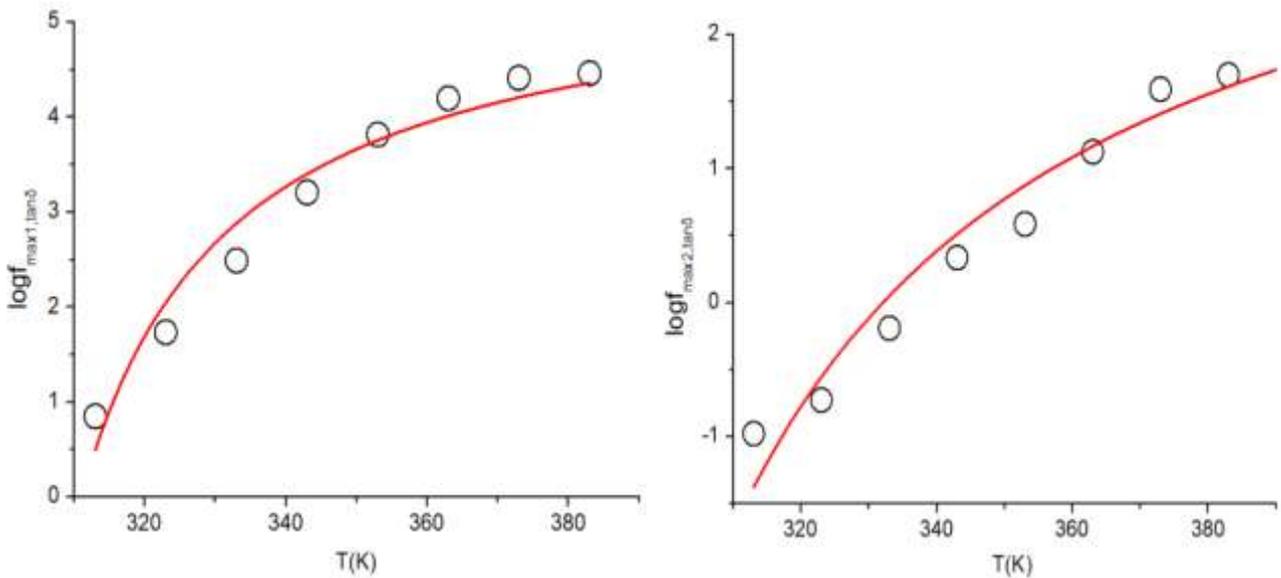

***Figure 10.*** *Best fits of eq. (3) to the data points. The fitting parameters are: $T_0$=289.9 K and $T_1$=118.3 K for DR1_tanδ and $T_0$=258.5 K and $T_1$=289.7K for DR2_tanδ.*

### *3.4.3 dc conductivity study through the electric modulus function*

$\sigma_{dc}$ can be obtained by fitting eq. (5) to the ε´´(f) data. In practice, this method can be, more or less, inaccurate due to undesirable low frequency electrode polarization. To overcome this problem, we used the complex electric modulus function: $M* = \frac{1}{\varepsilon*} = M' + iM''$, where:

$$M'(f) = M_s \frac{(f\tau_\sigma)^2}{1+(f\tau_\sigma)^2} \qquad (7)$$

$$M''(f) = M_s \frac{f\tau_\sigma}{1+(f\tau_\sigma)^2} \qquad (8)$$

A plot of $M''(f)$ is built up by a low frequency "conductivity peak", which results from the $\sigma_{dc}$ component itself, reaching a maximum at $f_{max,\sigma} = (2\pi\tau_\sigma)^{-1}$, or:

$$f_{max,\sigma} = \sigma_{dc}/(2\pi\varepsilon_0\varepsilon_\infty) \qquad (9)$$

where $\varepsilon_\infty \equiv \varepsilon'(f \to \infty)$. It is commonly assumed (and verified by our experimental data) that $\varepsilon'(f \to \infty)$ has a weak temperature dependence). Consequently, the FIT eq. (3) and its approximate one (eq.(4)) can be used to analyze the *f*$_{max, \sigma}$*(T)* data. We note that, dielectric relaxations are also recorded, but the maxima are shifted to higher frequency than those detected in ε´´(f) or *tanδ(f)* [25,26]. *M (f)* is a step-like plot; its high frequency plateau yields $M_s \equiv M'(f \to \infty)$ and the (relative) static dielectric constant $\varepsilon_s = \frac{1}{M_s}$, , as well. The electric modulus function, by definition, suppressed the undesirable low frequency space charge capacitance contributions, permitting a clear determination of both dc conductivity and static dielectric constant *ε*$_s$

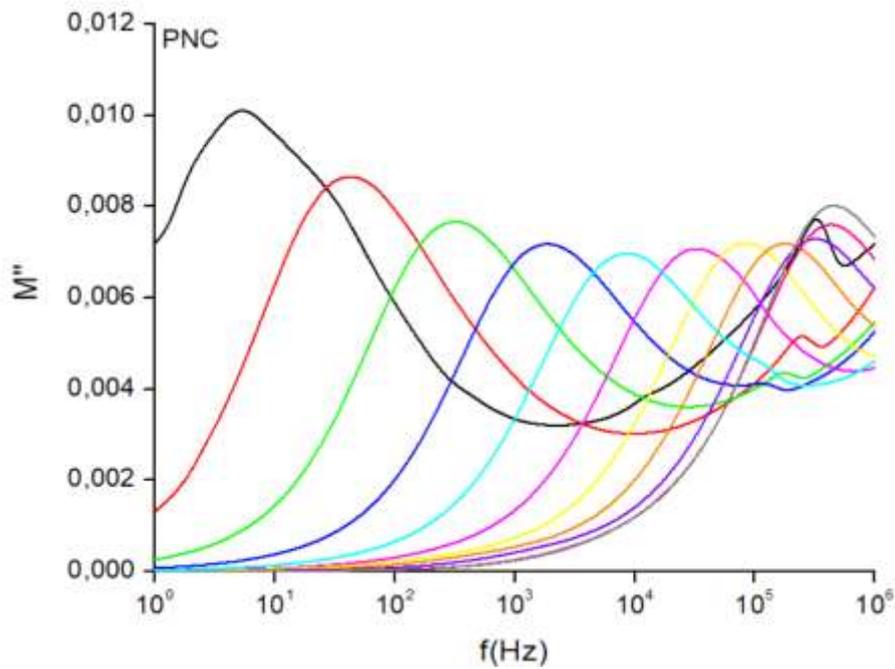

*Figure 11.* Variation of the imaginary part of electric modulus of PNC, recorded at different temperatures. The intense 'conductivity peak' shifts toweds higher frequenies, on increasing temperature.

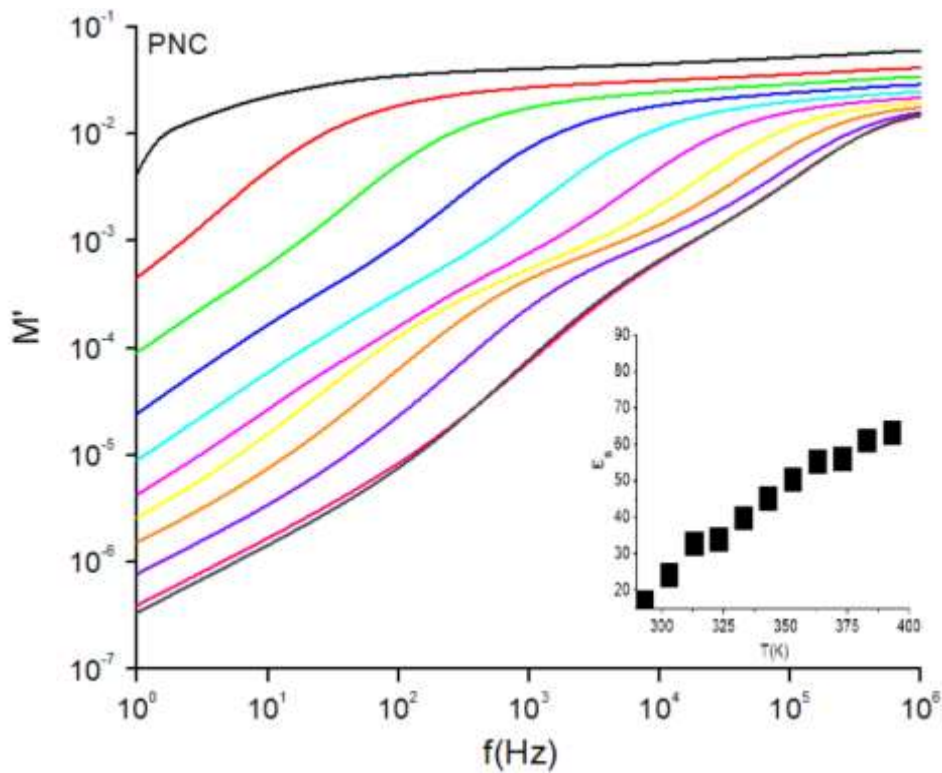

*Figure 12.* Isotherms of the real part of electric modulus of PNC, recorded on increasing temperatures ranging from 293 K (upper curve) to 393 K (bottom curve) .The relative static dielectric constant (inset diagram) was estimated from the limiting value $1/M'(f \rightarrow 10^6 Hz)$

The M´´(f) spectra is dominated by a low-frequency conductivity peak (Figure 11), while M´(f) tends to reach a saturation value in the high frequency limit (Figure 12). From the shift of the conductivity peak towards higher frequencies on increasing temperature (Figure 11), as explained above, the temperature evolution of $\sigma_{dc}$ is traced. In order to find out whether one or more dc conductivity mechanisms operate over the temperature range covered by BDS experiments, we checked if $logf_{max,\sigma}$ ($T^{-1/2}$) consists of one or more linear parts (each one corresponding to a different FIT mechanism described by eq. (2). indeed, in figure 13, the data points distribute over two successive linear regions, each one corresponding to an individual dc conductivity mode: C1 and C2, respectively. The use of the FIT model (eq. 1) provided the parameter $T_1$, from which, the activation energy $E \approx kT_1$ can be determined: 22(4) meV for C1 and 3(1) meV for C2, respectively. On increasing the temperature, the transition of the dc conductivity mechanism C1 to C2, signature a drop of the activation energy value by an order of magnitude and coincides with the onset of the kinetic process (attributed to the enhanced mobility of water molecules) detected in the first scan of the DSC thermograms. Dc conduction by FIT of electrons along the entire volume of the specimen capture thermal effects occurring in the polymer matrix.

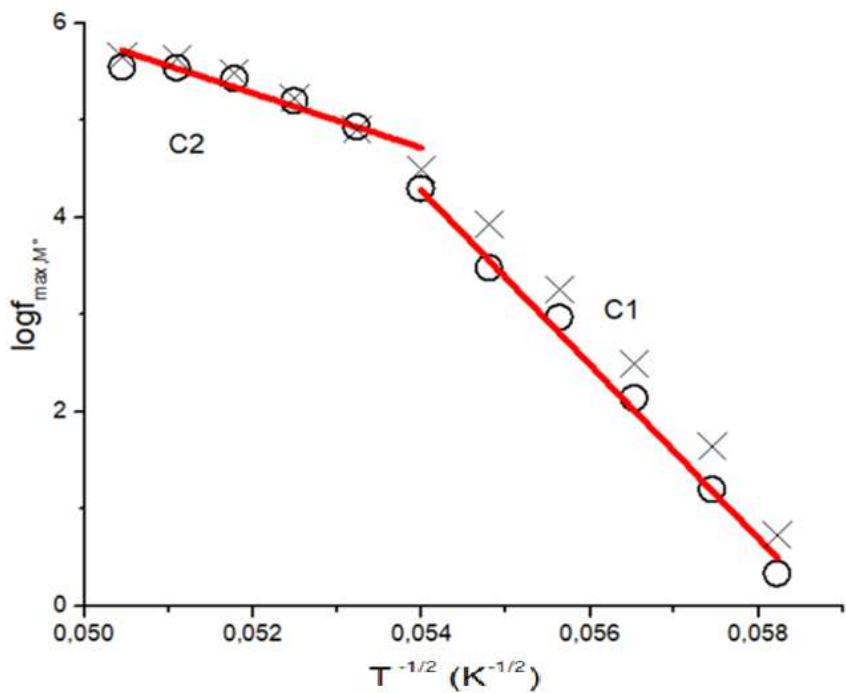

*Figure 13.* $T^{-1/2}$ *depedence of the conductivity peak observed in M´´(f). Each linear region signatures an inividual dc conductivity mechanism. The transition temperature compares with the initiation of the kinetic process recordes in th first scans of the DSC thermograms. Circles correspond to fresh PNCs and crosses to the piezo-electric one, respectively*

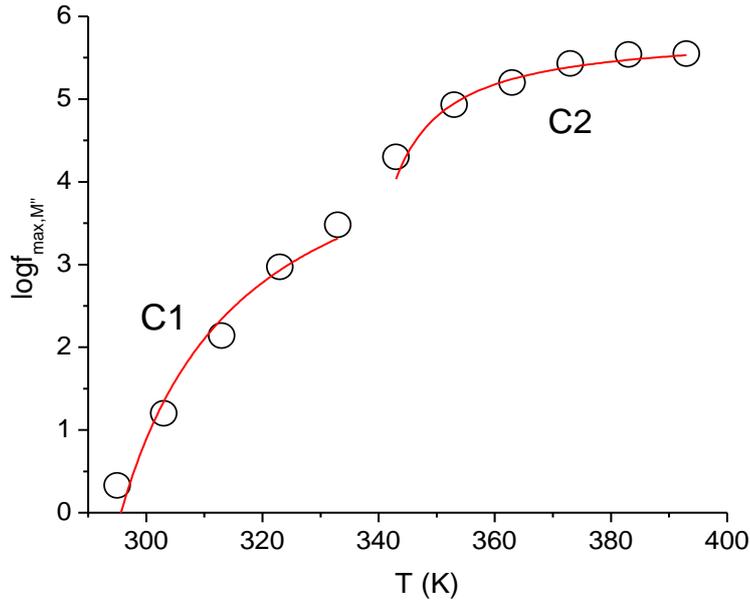

*Figure 14.* Lines are best fits of.th FIT model (eq. (2)) to the data points distributing in two successive temperature regions determined through the $T^{-1/2}$ representation (figure 13). The best fitting parameters of Eq. (4) are: $T_0=274.59K$ for C1 and $T_0=333.5K$ for C2.

## 4. Conclusions

Free standing blends of PVA/PVdF with dispersed NGPs were prepared by solution casting. Distinct electrical conductivity mechanisms, inspected through the electric modulus representation, operate in successive temperature regions. The temperature dependence of the dc conductivity obeys the fluctuation induced tunneling (FIT) model that describes inter-NGP electric tunneling. In comparison to DSC results, it was found that the dynamics of each conductivity mechanism coincides with structural phase transitions in the polymer matrix and the onset of a kinetic process related to an increased mobility traces of water molecules bound to PVA. A couple of dielectric relaxation mechanisms are related to the presence of NGPs in the polymer matrix and the temperature dependencies of their maxima evidence for (spatially localized) FIT of electrons within NGPs or aggregates. Fresh specimens render piezoelectric, after combined uniaxial stressing and electric polarization, with electromechanical coefficient comparable to neat piezoelectric PVdF. Both BDS and DSC experiments on piezoelectric composites showed that the electro-activity of PVdF grains has weak influence on the electric and thermal properties of the nano-composites. The electronic properties of the nano-composites studied in the present work can be tailored to match specific technological applications as sensors and switches, by tuning thermal and kinetic transitions of the polymer matrix

The harmonic electric field applied during a BDS sweep induces volume modifications of the electro-active PVdF grains and, subsequently, a time-varying mechanical stress field develops within the blend. On the other hand, electro-activity of PVdF grains can disturb the internal electric field that free (or bound) electric charges sense. Comparing the BDS spectra of as received and piezoelectric blends, we concluded that the effect of electro-activity of PVdF on the dc conductivity and dielectric relaxation is weak.